\documentclass[hyper]{JHEP3}

\input{epsf}
\usepackage{epsfig}
\usepackage{amssymb}
\usepackage{amsfonts}
\usepackage{amsbsy}

\def\IZ{{\mathbb{Z}}}

\def\IP{\mathbb{P}}

\def\CM {{\cal M}}

\def\CL {{\cal L}}

\def\CO {{\cal O}}

\def\CH {{\cal H}}

\def\half{\frac{1}{2}}

\newcommand{\eq}[1]{Eq.~(\ref{eq:#1})}

\def\one{{\hbox{ 1\kern-.8mm l}}}

\def\ba{\bar{a}}

\def\bz{\bar{z}}

\def\bz{\bar{z}}

\def\bra#1{{\langle}#1|}
\def\ket#1{|#1\rangle}

\def\vev#1{\langle{#1}\rangle}

\preprint{RUNHETC-2008-23}
\title{Black holes and balanced metrics}

\author{Michael R. Douglas$^{1,2,3}$ and Semyon Klevtsov$^{1,2,4}$\\
\\
$^1$  Simons Center for Geometry and Physics, 
Stony Brook University,\\
Stony Brook, NY 11794--3840, USA\\
\\
$^2$ NHETC and Department of Physics and Astronomy,
Rutgers University,\\
Piscataway, NJ 08855--0849, USA\\
\\
$^3$ I.H.E.S., Le Bois-Marie, Bures-sur-Yvette, 91440, France\\
\\
$^4$ ITEP, Moscow, 117259, Russia\\
\\
{\tt mrd@physics.rutgers.edu, klevtsov@physics.rutgers.edu} }

\abstract{
We consider a probe in a BPS black
hole in type II strings compactified on Calabi-Yau manifolds,
and conjecture that its moduli space metric is the balanced metric.
}

\begin{document}

\section{Introduction}

A famous problem in quantum gravity is to derive the
Bekenstein-Hawking entropy of a black hole by counting its
microstates.  In string theory, this was first done by Strominger and
Vafa \cite{Strominger:1996sh}.  They counted the microstates of a BPS
bound state of Dirichlet branes with the same charge as the black
hole, and then argued that the number of states was invariant under
varying the string coupling, turning the bound state into a black
hole.

This line of argument has been the basis for a great deal of work,
generalizing the result to other systems and away from the
semiclassical limit.  One important element in such results is the
claim that entropies and numbers of microstates are independent of the
moduli of the background.  An argument to this effect is provided by
the attractor mechanism \cite{Ferrara:1995ih}.  This was originally
stated for BPS black holes in type II strings compactified on a
Calabi-Yau manifold $X$, but the idea is probably more general 
(see \cite{Dabholkar:2006tb} for a recent discussion).  The attractor
mechanism is based on the observation that the equations of motion for
the moduli in a black hole background can be written in the form of
gradient flow equations for the area of a surface of fixed radius as a
function of the moduli.  This flow approaches an attracting fixed
point at the event horizon, with a definite value of the moduli and
area.  Thus, these values are insensitive to small variations of the
initial conditions.  By the Bekenstein-Hawking relation, this implies
that the entropy is invariant under such variations.

It is plausible that other properties of the black hole microsystem
share this type of universal behavior.  For example, we might
conjecture that not only the K\"ahler moduli of the Calabi-Yau metric
near a black hole take universal values, but that the entire metric is
universal, determined only by the charge and structure of the black
hole and independent of the asymptotic moduli.

What would this mean?  In classical supergravity, of course the metric
is determined by the Einstein equation, reducing to the Ricci flatness
condition for the source-free case.  Thus the stronger conjecture is
quite reasonable and indeed follows directly from the validity of
supergravity.  On the other hand, for a finite charge black hole
preserving eight or fewer supercharges, one knows that these equations
will get string theoretic ($\alpha'$ or $g_s$) corrections.  Thus,
while the stronger conjecture is still reasonable, it is not
{\it a priori} clear either what the attractor CY metric should be, or
what equations determine it.

Now, one reason the general question of finding exact metrics or
even precisely defining corrected
supergravity equations is hard, is that the metric and 
equations can be changed by field redefinitions, with no obvious
preferred definition.  For example, the metric $g_{ij}$ could be
redefined as $g_{ij} \rightarrow g_{ij} + \alpha R_{ij} + \beta
{(R^2)}_{ij} + \ldots$.  Unless we postulate an observable which singles
out one definition, say measurements done by a point-like observer who
moves on geodesics, there is no way to say which definition is right.
This problem shows up in computing $\alpha'$ corrections in the sigma
model as the familiar question of renormalization scheme dependence;
in general there is no preferred scheme.  We must
first answer this question, to give meaning to the ``CY attractor metric.''

A nice way to answer this question is to introduce a probe brane, say
a D0-brane, and study its world-volume theory.  The kinetic term for
its transverse coordinates is observable, and defines a unambiguous
metric on the target space, including any $\alpha'$ corrections.
While one can still make field redefinitions in the action, now these
are just coordinate transformations.
To make this argument straightforward, one requires that the mass (or
tension) of the probe be larger than any other quantities under
discussion, so that the action can be treated classically, and the
metric read off from simple measurements.\footnote{
This was the point of view taken in \cite{Douglas:1996yp,Douglas:1999ge}.
Actually, one can in principle reconstruct a manifold with metric
from quantum measurements (the spectrum and some position space 
observables), so one can work without this assumption.}
For example, this is true for D0-branes in
weakly coupled string theory, as their mass goes as $1/g_s$.  One can
then (in principle) define any term in the $g_s$ expansion this way.

Both on general grounds \cite{Douglas:1999ge} and in examples
\cite{Douglas:1997zj}, the moduli space metric seen by a D-brane probe
gets $\alpha'$ corrections, and for a finite size Calabi-Yau
background it is not Ricci flat.  The existing results are consistent
with the first such correction arising from the standard $\alpha'^3
R^4$ correction to supergravity \cite{Grisaru:1986px,Green:1981yb}, 
but pushing this to higher orders seems difficult.

Perhaps this problem becomes simpler in
a black hole background.  Rather than the D0,
the probe brane we will use is a D2 or M2-brane wrapped on the black hole
horizon.  As discussed in the works
\cite{Simons:2004nm,Gaiotto:2004pc,Gaiotto:2004ij,Aspinwall:2006yk},
such a brane, and D0-branes as well,
in a near horizon BPS black hole
background can preserve $SU(1,1|2)$ superconformal invariance.  This
is a symmetry of the $AdS_2\times S^2$ near horizon geometry and thus
this is as expected if multi-D0 quantum mechanics can be used as a dual
gauge theory of the black hole.  In these works, this quantum mechanics 
was argued to factorize into a space-time part, and an internal 
(Calabi-Yau) part; this second part describes motion of the probe
in the Calabi-Yau and can be used to define a probe metric.

Given this system and its relation to the black hole, we will give a
physical argument, based on the idea that a black hole must have
``maximal entropy'' no matter how this is defined, that suggests that
the probe metric in such a black hole background is in fact the
``balanced metric'' introduced and studied in the mathematics
literature
\cite{Tian:metrics,Donaldson1,Donaldson:numeric}.
As we explain, this is a condition which
determines a unique K\"ahler metric, in a way which {\it a
priori} is unrelated to the Ricci flatness condition.  
Nevertheless, one can show that in a certain large charge scaling limit
$k\rightarrow\infty$, the balanced metric approaches the Ricci flat
metric, with computable corrections in inverse powers in $k$.

Since our physical argument for the balanced metric does not assume the
equations of motion, it illustrates a way to
derive equations of motion from a maximum entropy principle. 
As pointed out to us by Vijay Balasubramanian, this idea was
suggested some time ago by Jacobson
\cite{Jacobson:1995ab,Eling:2006aw}
and might have more general application.

\section{BPS black holes and probes}

Let us consider a BPS black hole solution in IIa theory compactified
on a Calabi-Yau manifold $X$.  Such a solution is characterized by
discrete and continuous parameters.  The discrete parameters are its
electric and magnetic charges, which we take to be those of a system
of D0, D2 and D4-branes.  The continuous parameters are the values of
the hypermultiplet moduli, namely the dilaton, complex structure
moduli and their $N=2$ supersymmetry partners.  The vector multiplet
(K\"ahler) moduli are determined by the attractor mechanism, as we review
shortly.

As is well known, by varying the dilaton to strong coupling, this theory
is continuously connected to M theory compactified on $X\times S^1$.
In this theory, the black hole can be thought of as a black string
wrapped on $S^1$, and carrying $S^1$ momentum  \cite{Maldacena:1997de}.  
It will eventually
turn out that our conjecture appears more natural in M theory,
so let us start from that limit.  To get a black string, we can 
wrap M5-branes on a four-cycle
$[P]\in H_4(X, \IZ)$.  By Poincar\'e duality $[P]$ can also
be thought of as a class $p^A \omega_A$ in $H^2(X,\IZ)$, where
we introduce a basis $\omega_A$ of $H^2(X,\IZ)$. In general, there are also electric charges $q_A$,
corresponding to M2-branes wrapping dual two-cycles.
We will set these to zero in the subsequent discussion.

According to the attractor mechanism, the K\"ahler class $J_5$
of the CY at the horizon of the black string
is determined in terms of the charges $p^A$.
Unlike in $d=4$, in the black string solution, the volume $V$ of the CY is
a free parameter; thus we have (using 11d conventions of \cite{deBoer:2006vg})
$$
\frac{J_5}{V^{1/3}} = \frac{p^A \omega_A}{D^{1/3}} ,
$$
where $D \equiv D_{ABC} p^A p^B p^C$ and $D_{ABC}$ are the triple 
intersection numbers on the Calabi-Yau.  We recall that
$V=D_{ABC}J^A J^B J^C/6$ where $J=J^A\omega_A$.
It will also be useful to define 
\begin{equation}
\label{eq:scaleinv}
J_{CY}\equiv \frac {p^A \omega_A}{D^{1/3}}
\end{equation}
which is independent of the overall scale of the charges.  
The corresponding supergravity solutions simplify in the near-horizon
limit: the M theory solution approaches $AdS_3\times S^2$ geometry
\begin{equation}
\label{horizon}
ds^2 = L^2\left(\frac{-dt^2+dx_4^2+d\sigma^2}{\sigma^2} + d\Omega^2_{S^2}\right).
\end{equation}
with the following 4-form flux sourced by M5-branes 
$$
F_{(4)}\sim\frac1{L}\omega_{S^2}\wedge p^A\omega_A.
$$

Now, by compactifying the black string on $S^1$ with the radius $R_{10}$, we obtain a 4d black
hole with additional charge $q_0$, corresponding to momentum along the
string. At this level, the discussion is simply mapped into IIa
string theory, and the charges $(q_0,q_A,p^A)$
correspond to D0, D2 and D4 brane charges. The radius $L$ of $S^2$ and $AdS_3$  
is related to the radius $R_{10}$ of $S^1$ as $L\sim R_{10}\sqrt{D/q_0}$ and the volume of Calabi-Yau scales as $V\sim\alpha'^3q_0\sqrt{q_0/D}$. 
In $d=4$, the overall scale of $J$ and thus
the volume $V$ is also determined by the attractor mechanism, and \eq{scaleinv} becomes
\begin{equation}\label{eq:findJ}
J_4 = \alpha'\sqrt{\frac{q_0}{D}}p^A \omega_A,
\end{equation}
where as usual $\alpha'=l_p^3/R_{10}$ and $Q$ is the graviphoton charge (10d conventions correspond to those of \cite{Gaiotto:2004ij}).
The IIa supergravity solution in four dimensions approaches the $AdS_2\times S^2$ near horizon geometry.

The metric on the CY $X$ is determined by the Einstein equations,
\begin{equation}\label{eq:ein}
R_{ij} - \half R g_{ij} \sim (F^2)_{ij} + \delta + \CO(l_p)
\end{equation}
where $(F^2)_{ij}$ schematically
denotes the contribution to stress-energy density due to the RR fields
of the charged black hole, and $\delta$ denotes the contribution from brane sources \cite{Bachas:1999um}.
The last term on the rhs represents string or 
M-theoretic corrections to this equation \cite{Green:1981yb,Green:1997di}.

In general, this metric depends on details of the state of the black
hole.  For example, if there are localized brane sources, the metric
will depend on their location. However, we can simplify it by
considering an appropriate state of the black hole. 
For general $X$, despite the non-zero rhs in \eq{ein},
there exists a black hole state in which the sources are averaged in
an analogous way, removing the localized terms, and leading to a Ricci
flat metric on $X$ in the supergravity limit.  

At first it may seem that the field
strength $F$ would lead to a complicated position-dependent source;
say concentrated on the cycles wrapped by the M5-branes.
However, as observed in \cite{Aspinwall:2006yk}, the combination
of the attractor mechanism and the equations of motion force the
field strength to be
proportional to the K\"ahler form on $X$,
\begin{equation}\label{eq:propto}
F = dA =p^A\omega_A= Q J,
\end{equation}
where $Q$ is the graviphoton charge $Q\sim\sqrt{D/q_0}$.
This is true before adding $\alpha'$ or $g_s$ corrections, and all the
simple candidate corrections one can write down (such as powers of $F$
and its derivatives, for example those which appear in the MMMS
equation \cite{Marino:1999af}), preserve \eq{propto}. 
While there are still $F^2$ sources in \eq{ein}, one can now check
that these are canceled by terms coming from the space-time dependence
of radius $R_{10}$ of 11th dimension (in M theory) or the dilaton (in IIa).  We omit the
explicit check, instead pointing out that since the sources are
constructed only from the metric tensor on $X$, they can at most add a
cosmological constant term in \eq{ein}, leading to a K\"ahler-Einstein
metric.  However, standard arguments imply that a solution exists only
if this term is proportional to $c_1(X)$, which is zero for a CY.

We have thus defined a preferred state of the black hole, for which
the metric on $X$ would be Ricci flat in the supergravity limit. To
study corrections, we need to define this preferred state in a way
which does not assume that we know the correction equations of motion
or their solution in advance.  Since the supergravity argument
required us to average over all of the internal structure of the black
hole, it is natural to define it as a mixed state of maximal
entropy, as we will do below.

\section{The probe theory}
\label{s:probe}

Having defined the state of the black hole under consideration, we now
proceed to define the metric with $l_p$ (11d Planck length) or $\alpha'$
corrections.  As we discussed earlier, this can be made precise by
introducing a probe, whose moduli space (space of zero energy
configurations) includes $X$.  This typically requires that the probe
preserves some supersymmetry.

Now, the BPS black hole preserves an $N=1$ supersymmetry, determined  by
the phase of its central charge $Z$, which is determined by the
charges and attractor moduli.  In asymptotically Minkowski space-time,
introducing another BPS brane will typically break all of the
supersymmetry.  However, it was shown  in \cite{Simons:2004nm}
that in the near-horizon limit,
a probe zerobrane can nevertheless preserve space-time supersymmetry,
if it follows its ``charged geodesic'' ({\it i.e.} trajectory determined
by the background metric and RR field).  Even a collection of such branes
with misaligned charges can preserve supersymmetry; consistent with this,
the combined gravitational and RR potential energy of such a collection
is additive.

Choosing a probe brane which preserves supersymmetry, one expects its
configuration space to be some moduli space associated with the
compactification space $X$.  In the simplest example of a D0-brane,
the moduli space is $X$ itself.  Another example for which the moduli
space is $X$ is a D2-brane wrapped around the $S^2$ horizon.  Other
choices, for example a D$p$-brane wrapped on a $p$-cycle of $X$, would
lead to different moduli spaces, related to the geometry of $X$.

While there will be a probe world-volume potential, this is determined
by superconformal invariance \cite{Gaiotto:2004pc} to be a function of
the radius $\sigma$, but independent of the other coordinates.  In
particular, it is independent of position on $X$ or other ``internal''
coordinates.  Thus the world-volume theory includes a supersymmetric
quantum mechanics on the moduli space $X$.  Ground states of this
quantum mechanics will correspond in the usual way to differential
forms on $X$.

The argument we are about to make is clearest for the case of a probe
D2 wrapping the $S^2$ horizon, so let us consider that.  
According to \cite{Gaiotto:2004pc}, the supersymmetry condition for
such a brane forces it to the center of $AdS_2$ (in global coordinates),
so there are no other moduli on which the probe metric on $X$ can depend.

At leading order, the probe will see both the metric on $X$, and a
magnetic field on $X$.  The latter follows (in IIa language) from the
D4 charge of the black hole: a probe
D2 wrapping the horizon will see a background magnetic field on the
CY $X$ \cite{Gaiotto:2004ij,Aspinwall:2006yk}, 
$$
F_{CY} = \int_{S^2} F_{(4)} = p^A \omega_A .
$$
Mathematically,
such a magnetic field defines a line bundle $\CL$ over $X$;
whose first Chern class is the D4 charge $p^A$.
From \eq{findJ}, the K\"ahler
class $J$ is proportional to the first Chern class of $\CL$,
$$
J = \frac{1}{Q} c_1(\CL) .
$$

\section{Maximal entropy argument for the probe metric}

Our argument will be based on two assumptions.  First, the most
symmetric state of a BPS black hole, and thus the state corresponding
to the simplest metric on $X$, is a state of maximal entropy.
Second, that there is a sense in which the black hole can be
regarded as made up of constituents with ``the same dynamics'' as
the probe.  We will use this to argue that the probe should
also be in a state of maximal entropy to get a simple result.

The first assumption is very natural and straightforward to explain.
To define ``maximal entropy,'' we look at the Hilbert space of BPS
states of the black hole, call this
$\CH_{BH}$.  By standard arguments going back to
\cite{Strominger:1996sh}, these are BPS states of the quantum
system describing the black hole, here a bound state of D0 and D4
branes.  Let us denote an 
orthonormal basis of $\CH_{BH}$ as $\ket{h_\alpha}$.
Now, the states $\ket{h_\alpha}$ 
are pure states in the usual sense of quantum mechanics.
The maximal entropy state of such a system is a mixed state, described
by the density matrix
\begin{equation}\label{eq:bh-dens}
P_{BH} = \frac{1}{\dim\CH_{BH}}\sum_\alpha \ket{h_\alpha}\bra{h_\alpha} ,
\end{equation}
in which each pure state appears with equal probability.
Thus, we have a clear definition of ``maximal entropy'' of the black hole.

The original description of the black hole Hilbert space $\CH_{BH}$
\cite{Balasubramanian:1996rx}
was in terms of a postulated bound state of D0-branes at each triple
intersection of D4-brane on the Calabi-Yau. Denoting the number of triple
intersections as $k$, one finds that the supergravity entropy formula
can be matched if there is one D0 bound state for each value $n$ of 
eleven-dimensional momentum, with $4$ bosonic and $4$ fermionic
degrees of freedom, leading to a partition function
\begin{equation}\label{eq:partition}
Z_{BH} = \prod_{i=1}^k \prod_{n\ge 1} \left(\frac{(1+q^n)}{(1-q^n)}\right)^4 
\end{equation}

A later argument to the same effect
\cite{Maldacena:1997de} proceeds by
lifting the black hole to M theory on $X \times S^1$, in which
it becomes a wrapped M5-brane.  First compactifying on $X$, 
a wrapped five-brane on a 4-cycle (or divisor) $D$ becomes a black string.
The string is then compactified on $S^1$ to obtain the black hole.

In this analysis, the string has world-sheet fields parameterizing the
moduli space of degree $N$ hypersurfaces $P_N$, which is precisely 
the projectivization of the space of sections of $\CL$.
The resulting black string Hilbert space is that of a symmetrized
orbifold ${\rm Sym}^M(\CM(P_N))$ of the moduli space, constructed as
\begin{equation} \label{eq:bhstates}
\prod_{i=1}^k \alpha_{-n_i}^{A_i} \ket{0}
\end{equation}
with $\sum n_i=M$.  Along with these moduli are additional fields 
(the dimensional reduction of the fivebrane two-form, and fermions),
combining into $(0,4)$ supersymmetry multiplets with $4+4$ components.
Finally, using the standard result for the density of states of a 
conformal field theory with central charge $c=6k$,
the entropy is
$$
S = 2\pi\sqrt{\frac{c\cdot q_0}{6}} .
$$
Another, more mathematical way to derive the multiplicity $4+4$ for
each modulus, is to observe that the BPS states of
$(0,4)$ supersymmetric quantum mechanics include
all $(p,q)$ forms taking values in the target space (here the 
moduli space of divisors),
\begin{equation}\label{eq:defH1}
\CH = \oplus_{0\le p,q\le 3} H^p(X,\Omega^q \otimes \CL).
\end{equation}
Since the divisor is ample, these vanish for $p>0$, while the
$q=0,1,2,3$ terms have multiplicities $k,3k,3k,k$ (for large $D$).
The even and odd $q$-forms then give rise to bosonic and fermionic
moduli (respectively), whose quantization reproduces \eq{partition}
or \eq{bhstates}.

More recently, a related but not obviously identical
description of the black hole Hilbert space has been developed,
motivated by the idea that the black hole should be described by
a superconformal matrix quantum mechanics of $n$
D0-branes in the D4 background.
\cite{Gaiotto:2004ij,Aspinwall:2006yk}
In this picture, the basic object is a bound state of $n$
D0-branes which can be thought of as a ``fuzzy D2-brane,'' which 
arises from the matrix D0 theory by a Myers-type effect
\cite{Myers:1999ps}.  The general form of \eq{partition} then
arises by summing over all partitions of the 
total D0 charge $q_0$.

Note that this second description is in terms of a supersymmetric
quantum mechanics with target space the Calabi-Yau manifold $X$, very
much like our probe theory.  Indeed, the background RR field is
postulated to appear as a non-trivial $U(1)$ magnetic field, of
topological type exactly that of the bundle $\CL$.  

A strategy to get the D0 matrix quantum mechanics on this background,
pursued in \cite{Gaiotto:2004ij}, is to consider a D2-brane
wrapped on the black hole horizon, an $S^2$.  As is familiar
(for example) in M(atrix) theory \cite{deWit:1988ig,Banks:1996vh}, 
D0 matrix quantum
mechanics contains bound configurations of $N$ D0's which represent
a wrapped or stretched D2.  If we can reverse this identification, we
can derive the matrix quantum mechanics from the D2 theory.

Of course, the D2 theory in this background is precisely the probe theory
we discussed in section \ref{s:probe}.
Its full low-energy hamiltonian was found in
\cite{Gaiotto:2004ij}.  It factorizes into an $AdS_2$ part and
a $CY$ part, with the latter being
\begin{equation}\label{eq:cy-qm}
H_{CY}=g^{a\ba}(P_a-A_a)(P_{\ba}-A_{\ba}).
\end{equation}
Here the metric $g$ is built from the K\"ahler form $J$ and the
gauge field has the field strength proportional to $J$ as in
\eq{propto}.
The general idea is then that,
by promoting this quantum mechanics to matrix quantum mechanics, one
would obtain a description of the black hole.

This brings us to our second assumption, that there is a sense in
which the black hole can be regarded as made up of constituents with
``the same dynamics'' as the probe.  If we grant the second description
of the black hole, in terms of D0 matrix quantum mechanics,
then clearly we can identify constituents with the
same dynamics as our probe.  As we mentioned, reproducing the black
hole partition function \eq{partition} requires summing over configurations
each labelled by a partition $\{n_i\}$ of the total D0 charge.  
Such a configuration is obtained by considering the matrix variables
as a direct sum of blocks, each a matrix of dimension $n_i$.  The dynamics
of such a block  is described by the $U(n_i)$ reduction
of the matrix quantum mechanics, with interactions with the rest of
the black hole produced by integrating out off-diagonal degrees of
freedom.  The supersymmetry of the combined system will cancel the
relative potential between the blocks, and presumably makes the other
induced interactions small.

Let us consider a sector with $n_1=1$, in other words containing single 
unbound D0.  The dynamics of this D0 is approximately described by the
$U(1)$ version of matrix quantum mechanics, in other words the theory
discussed in \cite{Gaiotto:2004ij}.   The BPS Hilbert space of this theory
is $\CH$ defined in \eq{defH1}, and we see that this sits naturally in
$\CH_{BH}$.

Now, we will implement our second assumption, by deriving a natural
maximal entropy state for the probe.  By starting with the maximal
entropy state \eq{bh-dens} of the black hole in $\CH_{BH}$ and tracing over
all of the other degrees of freedom, we obtain a density matrix $P$
over the Hilbert space $\CH$.  The result will be the standard
quantum state of maximal entropy for this quantum mechanics, which
assigns equal probability to each state in $\CH$, given by
the expression 
\begin{equation}\label{eq:dens}
P = \frac{1}{\dim\CH}\sum_\alpha \ket{h_\alpha}\bra{h_\alpha} ,
\end{equation}
Indeed, one might regard this choice of quantum state as the natural
one whatever the probe is, without calling upon any relation to the
black hole.  However we spell out this step as it explains how we 
could, given a precise D0 quantum mechanics for the black hole, compute
the probe state and observables.

Now, given $P$, we can ask, what is
the probability to find the D2 probe at a given point $z\in X$.
This will be
\begin{equation}\label{eq:prob}
P(z,\bz) = \bra{z} P \ket{z} .
\end{equation}
In general, the D2 will have ``spin'' degrees of freedom as well,
corresponding to the degrees $(p,q)$ of cohomology; let us fix these
in the $p=q=0$ sector.\footnote{We comment on this
point in section \ref{conclusion}.}  By inserting explicit
wave functions $\psi_\alpha(z,\bar z)$,
the density matrix can be written in position space as a kernel,
\begin{equation}\label{eq:kern}
P(z_1,\bz_1,z_2,\bz_2) =
 \frac{1}{\dim H^0}
 \sum_\alpha \psi^*_\alpha(z_1,\bz_1)~\psi_\alpha(z_2,\bz_2) ,
\end{equation}
with \eq{prob} its values on the diagonal $z_1=z_2=z$.

Note that, although the lowest Landau level wavefunctions satisfy the
metric-independent linear differential equation $\bar D h=0$, their
normalizations depend on the metric.  Thus the kernel \eq{kern}
depends on the specific choice of metric, not just the K\"ahler class.

Now, since the probe has maximal entropy, one would expect
that this probability does not favor any particular point in moduli
space, in other words
\begin{equation}\label{eq:constP}
P(z,\bz) = {\rm constant} .
\end{equation}
But this is not at all obvious from what we have said so far.
We might regard it as a second, independent interpretation of the
claim that the black hole has maximal entropy.

While from the point of view of an asymptotic observer, the first
definition \eq{dens} of maximal entropy seems more natural, if we can
only make measurements with the probe, the second definition seems
more natural.  Going further, to the extent that (following the
arguments above) the probe can also be thought of as a constituent of
the black hole, we might be able to reformulate black hole
thermodynamics in terms of the second definition.  In particular, the
postulate that the black hole has maximal entropy, should imply that
its constituents are equidistributed in moduli space. Otherwise, there
would be a simple way for the system to increase its entropy

To summarize, while not self-evident, it is an attractive hypothesis
that the entropy should be maximal in both senses.  Actually, the two
definitions of maximal entropy are not directly in conflict.  Indeed,
we could compute \eq{prob} from the definition \eq{kern}, and check
whether they agree.  But since the actual wave functions and thus
\eq{kern} depend on the details of the probe world-volume theory, in
particular the metric, we need to know the probe metric to make this
check.

Turning around this logic, we can regard the conjunction of \eq{dens}
and \eq{constP} as a non-trivial condition on the probe metric.  In
fact, this is a known condition: it implies that the
probe metric is the balanced metric.

\section{Balanced metrics}

We begin by recalling the relation between a ``physical'' wave function
$\psi(z,\bar z)$, for which the inner product is
\begin{equation}\label{eq:ip}
\vev{\psi|\psi'} = \int_M d{\rm vol}\ \psi^*(z,\bar z) \psi'(z,\bar z) ,
\end{equation}
and a holomorphic section $s(z)$ of a line bundle.  We assume that $\psi$
couples minimally to a $U(1)$
vector potential $A_a$, so that the Hamiltonian is written
in terms of covariant derivatives
\begin{eqnarray}
D_a  = \frac{\partial}{\partial z^a} + iA_a \\
{\bar D}_{\bar a}  = \frac{\partial}{\partial {\bar z}^{\bar a}}
  - iA_{\bar a} .
\end{eqnarray}
Now, if $0 =F^{0,2}=\bar\partial \bar A$, there will exist
a complex gauge transformation $g(z,\bar z)$ such that
$$
g^{-1}\, {\bar D}_{\bar a}\, g = {\bar\partial}_{\bar a} .
$$
Defining
\begin{equation}\label{eq:gt}
\psi = g \cdot s ,
\end{equation}
we convert the condition $\bar D\,\psi=0$ to the holomorphy condition
${\bar\partial}\,s=0$, at the cost of turning the inner product \eq{ip}
into
\begin{equation}\label{eq:mathip}
\vev{s|s'} = \int_M d{\rm vol}\ e^{-K}\, s^*(\bar z) s'(z) 
\end{equation}
with
\begin{equation}\label{eq:defK}
e^{-K(z,\bar z)} \equiv |g(z,\bar z)|^2 .
\end{equation}
Mathematically, the factor $e^{-K}$ is referred to as a
hermitian fiber metric on $\CL$; it allows one to
multiply two sections $s$ and $s'$ pointwise to get a function.

Physically, this is useful as a minimal energy (or supersymmetric)
gauge field will satisfy $F^{0,2}=0$, while ground state wave functions
will satisfy ${\bar D}\psi=0$ (the lowest Landau level, 
here BPS states).\footnote{
As is well known, these states can also be used to 
define ``fuzzy'' or noncommutative versions of a space.
This was tried in the present context in \cite{Gaiotto:2004pc};
there may be connections to \cite{Iqbal:2003ds}, or to other 
appearances of noncommutative geometry in string and M theory.}

One reason to make the definition \eq{defK} is that
the condition \eq{propto} becomes
$$
J\propto F^{(1,1)} = \partial{\bar\partial}K ,
$$
in other words $K$ is proportional to the
K\"ahler potential on $X$.

Now, let us recall the definition of the balanced metric from 
\cite{Tian:metrics,Donaldson1,Donaldson:numeric} (see also \cite{DKLR,Douglas:2008pz} for discussions with more physical introduction).
An ample line bundle $\CL$ over $X$ defines an embedding of $X$
into $\IP(H^0(X,\CL))$, the projective space parameterized by a basis
of sections of $\CL$.    Let us choose a basis $s^\alpha$ of 
these sections, and define it by fiat to be orthonormal.
Having done this, we have a Fubini-Study metric $J_{FS}$ on 
this projective space, with
K\"ahler potential given by the standard expression in terms
of the homogeneous coordinates $s^\alpha$,
\begin{equation}\label{eq:FSK}
K=\log \sum_\alpha |s_\alpha|^2 .
\end{equation}
We can then pull this back to $X$ to
get a K\"ahler metric $J_X$ on $X$ (formally
with the same K\"ahler potential, but evaluated on the subset $X$).

As in \eq{mathip}, this metric can then be used to define an inner
product on sections,
\begin{equation}\label{eq:secip}
\vev{s'|s} = \int_X d{\rm vol}(J_X)\ %
 \frac{(s')^*~ s}{\sum_{\alpha} |s^\alpha|^2} ,
\end{equation}
where ${\rm vol}(J_X)=J_X^n/n!$ is just the usual volume
form $\sqrt{g}$ for this K\"ahler metric.

Now, there are two senses of an ``orthonormal basis of sections,''
the one we introduced in the beginning by fiat, and a second one
defined by \eq{secip}.  In general, these will not agree.  But,
if they do, we refer to this as a ``balanced'' embedding, and
the resulting metric $J_X$ as the balanced metric.

Now, going back to \eq{kern}, this is a squared sum over a
basis of orthogonal wave functions in the usual physical inner product
\eq{ip}.  However, tracing through the definitions, \eq{secip} is
the same as \eq{ip}, just rewritten using \eq{gt} and the
definition of the pull-back metric.  Now, the balanced condition
implies that this orthonormal basis, is the {\it same} as the
orthonormal basis $s^\alpha$ we chose at the start.  

Thus, using \eq{FSK}, we have
\begin{eqnarray}
\label{eq:balance}
P(z,\bz) &=& \frac{\sum_\alpha s_\alpha(z) {\bar s}_\alpha(\bz')}
 {\sum_{\alpha} |s^\alpha(z)|^2}\bigg|_{z'=z} =1
\end{eqnarray}
so indeed the kernel, and thus the probability distribution for
the probe brane, is constant precisely for the balanced metric.

Conversely, it is a theorem that (given suitable assumptions, which
hold here), the balanced metric exists and is unique
\cite{Donaldson1}; thus this is the only metric satisfying both
\eq{constP} and \eq{propto}.
Thus, granting \eq{propto},
our physical consistency condition between the two definitions
of ``maximal entropy,'' precisely picks out the balanced metric
associated to the line bundle $\CL$ whose first Chern class is the
D4 charge.

Now, in the limit of a large charge black hole, in which the local
curvatures and field strengths near the black hole become small, one
would expect the probe metric to be approximately Ricci flat.  To take
the large charge limit, we scale up the D4 charge by a factor $k$, and
take $A\rightarrow k\, A$.  Mathematically, this corresponds to
replacing the line bundle $\CL$ by the line bundle $\CL^k$ (defined as
a tensor product).

The claim is now that, in the large $k$ limit, the balanced metric,
defined by the maximal entropy property \eq{balance}, should satisfy
the supergravity equations of motion. By the discussion following
\eq{ein}, these equation imply that the metric $X$ will be Ricci flat.
But {\it a priori}, the condition \eq{balance} has no evident
connection with Ricci flatness or any other equation of motion.
Thus this claim is in fact a nontrivial test of the conjecture.
And it passes this test, as we explain in the next section.

Let us finally comment on the appropriate large $k$ scaling limit in
terms of black holes charges. Looking at \eq{scaleinv} one sees that
in M-theory settings the above scaling corresponds just to rescaling
of the magnetic charges $p^A\rightarrow k\,p^A$.  In the IIa set up,
\eq{findJ} and \eq{propto} tell us that in addition to rescaling the
magnetic charges $p\rightarrow kp$, one also has to scale
$q_0\rightarrow kq_0$, so that the curvature of line bundle $\CL^k$
scales as $k$ times the metric. Such a scaling limit is described in
\cite{Denef:2007vg} as the natural limit scaling up K\"ahler moduli.

\section{Comparisons and conclusions}
\label{conclusion}

As it turns out, we can get precise results for the limit
$k\rightarrow\infty$ of the density matrix \eq{dens}, using an
asymptotic expansion for the diagonal of the kernel \eq{kern}
developed in \cite{Tian:metrics,Zelditch:Szego,Catlin,LuZhiqin}.  We
refer to \cite{Douglas:2008pz} for a physical explanation and derivation of
this expansion, and merely cite the result here:
\begin{equation}
\label{eq:expan}
P(z,\bz) = \frac{k^3}{\dim H^0}\left(1 + \frac{1}{k} R + 
\frac1{k^2}\left(\frac13\Delta R +
 \frac1{24}(|R_{a\bar ab\bar b}|^2-4|R_{a\bar a}|^2+3R^2)\right) +
\ldots\right) .
\end{equation}
where $R, R_{a\bar a}$ and $R_{a\bar ab\bar b}$ are the scalar
curvature, Ricci and Riemann tensors of the metric whose K\"ahler form
is the curvature of $\CL$, {\it i.e.} $J=F$.

The expansion (\ref{eq:expan}) is a general statement, but combining it with 
\eq{balance} implies that, for sufficiently large $k$, the balanced metric will
have constant scalar curvature, up to corrections of order $1/k$.
In the case at hand with $c_1(X)=0$, this implies Ricci flatness, and
the probe metric satisfies this test at leading order.  Thus the basic
consistency of our conjecture with supergravity is clear.

Another variation of the conjecture is that, because the probe is a
superparticle, we should also sum over spin states in the
density matrix \eq{kern}, and enforce \eq{balance} on the diagonal of
this density matrix.  In \cite{Douglas:2008pz}, we show how to define and
compute the leading terms of this kernel using supersymmetric quantum 
mechanics.  The resulting expansion is very similar.  For example,
the leading nontrivial term, proportional to the Ricci scalar,
now comes with a coefficient $(1-N)$, where $N$ is the number of 
supercharges.  

In the case at hand, the probe is an $N=2$ supersymmetric quantum mechanics.
Thus incorporating the spin states leads to the same basic result,
that in the large volume limit the conjecture agrees with supergravity.

However, we have not been able to identify the subleading terms in
\eq{expan}, nor those in its supersymmetric analogs, with any known
physical corrections.  In particular, one might expect the famous
coefficient $\zeta(3)$ of the $R^4$ correction to show up in this
expansion, from both the IIA string and M theory points of view.  On
the other hand, it is clear from the nature of the expansion
\eq{expan} that such transcendental coefficients will not appear.

It seems possible that the limits involved kill these particular
terms, but one still needs to explain the other corrections.  One
might speculate that these are related to the much studied higher
genus superpotential terms in RR backgrounds
\cite{Neitzke:2004ni,Marino:2004eq}, but we did not find evidence for
this either.  Or, it could be that the problem is essentially
nonperturbative from the supergravity point of view, and that these
terms are seeing another regime.

One might ask if the expression \eq{propto} for the magnetic field
could also get stringy or M-theoretic corrections, which while
preserving its cohomology class, nevertheless modify \eq{expan}.  This
is not possible because, as also shown in \cite{Douglas:2008pz}, the
degeneracy of the lowest Landau level for the Hamiltonian \eq{cy-qm}
requires the gauge connection to satisfy the hermitian Yang-Mills
equation, which for a $U(1)$ gauge field implies \eq{propto}.

Another possibility is that additional couplings on the probe
world-volume are important.  We assumed that the probe can be
described purely in terms of a particle in a background metric and
magnetic field, and then derived the balanced metric from the
maximal entropy condition.  Of course, there could be higher derivative
terms, perhaps induced by interactions with the other constituents
of the black hole.  These might modify the wave functions so as to
achieve \eq{balance} with a different metric.

Of course, the maximal entropy assumptions might not hold for these
black holes.  We nevertheless feel that our argument is making an
important point.  The assumptions do seem very natural in the context
of this problem.  Indeed, if we had a precise definition of the D0
matrix quantum mechanics suggested in
\cite{Gaiotto:2004ij,Aspinwall:2006yk}, we could in principle use it
to compute the probe metric, and find out where the contradiction
arises.  Indeed, understanding this point might be a useful hint to
how this (still mysterious) quantum mechanics works.

Furthermore our argument is very simple, indeed far simpler than the
supergravity or topological string considerations one might compare it
with.  We believe it will find applications regardless of the fate of
this conjecture.

{\bf Acknowledgments} We thank V. Balasubramanian, J. de Boer,
F. Denef, S. Donaldson, R. Karp, J. Keller, S. Lukic, G. Moore ,
N. Nekrasov, B. Pioline, R. Reinbacher, M. Ro{\v c}ek, D. Sahakyan,
G. Torroba, D. Van den Bleeken and P. Vanhove for valuable
discussions. This work was supported in part by DOE grant
DE-FG02-96ER40959. The work of S.K. was also supported by the grant
for support of scientific schools NSh-3035.2008.2, RFBR grant
07-02-00878 and by the Department of Physics and Astronomy at
Rutgers. S.K. thanks members of YITP at Stony Brook for their warm
hospitality.


\begin{thebibliography}{99}

%\cite{Strominger:1996sh}
\bibitem{Strominger:1996sh}
  A.~Strominger and C.~Vafa,
  ``Microscopic Origin of the Bekenstein-Hawking Entropy,''
  Phys.\ Lett.\ B {\bf 379}, 99 (1996)
  [arXiv:hep-th/9601029].
  %%CITATION = HEP-TH 9601029;%%


%\cite{Ferrara:1995ih}
\bibitem{Ferrara:1995ih}
  S.~Ferrara, R.~Kallosh and A.~Strominger,
  ``N=2 extremal black holes,''
  Phys.\ Rev.\ D {\bf 52}, 5412 (1995)
  [arXiv:hep-th/9508072].
  %%CITATION = HEP-TH 9508072;%%

\bibitem{Dabholkar:2006tb}
  A.~Dabholkar, A.~Sen and S.~P.~Trivedi,
  ``Black hole microstates and attractor without supersymmetry,''
  JHEP {\bf 0701}, 096 (2007)
  [arXiv:hep-th/0611143].

%\cite{Douglas:1996yp}
\bibitem{Douglas:1996yp}
  M.~R.~Douglas, D.~Kabat, P.~Pouliot and S.~H.~Shenker,
  ``D-branes and short distances in string theory,''
  Nucl.\ Phys.\  B {\bf 485}, 85 (1997)
  [arXiv:hep-th/9608024].

%\cite{Douglas:1999ge}
\bibitem{Douglas:1999ge}
  M.~R.~Douglas,
  ``Two lectures on D-geometry and noncommutative geometry,''
  [arXiv:hep-th/9901146].
  %%CITATION = HEP-TH 9901146;%%

\bibitem{Douglas:1997zj}
  M.~R.~Douglas and B.~R.~Greene,
  ``Metrics on D-brane orbifolds,''
  Adv.\ Theor.\ Math.\ Phys.\  {\bf 1}, 184 (1998)
  [arXiv:hep-th/9707214].

%\cite{Grisaru:1986px}
\bibitem{Grisaru:1986px}
  M.~T.~Grisaru, A.~E.~M.~van de Ven and D.~Zanon,
  ``Four Loop Beta Function For The N=1 And N=2 Supersymmetric Nonlinear Sigma
  Model In Two-Dimensions,''
  Phys.\ Lett.\  B {\bf 173}, 423 (1986).

%\cite{Green:1981yb}
\bibitem{Green:1981yb}
  M.~B.~Green and J.~H.~Schwarz,
  ``Supersymmetrical String Theories,''
  Phys.\ Lett.\  B {\bf 109}, 444 (1982).
  %%CITATION = PHLTA,B109,444;%%
  
%\cite{Simons:2004nm}
\bibitem{Simons:2004nm}
  A.~Simons, A.~Strominger, D.~M.~Thompson and X.~Yin,
  ``Supersymmetric branes in AdS(2) x S**2 x CY(3),''
  Phys.\ Rev.\ D {\bf 71}, 066008 (2005)
  [arXiv:hep-th/0406121].
  
%\cite{Gaiotto:2004pc}
\bibitem{Gaiotto:2004pc}
  D.~Gaiotto, A.~Simons, A.~Strominger and X.~Yin,
  ``D0-branes in black hole attractors,''
  [arXiv:hep-th/0412179].
  %%CITATION = HEP-TH 0412179;%%
  
%\cite{Gaiotto:2004ij}
\bibitem{Gaiotto:2004ij}
  D.~Gaiotto, A.~Strominger and X.~Yin,
  ``Superconformal black hole quantum mechanics,''
  JHEP {\bf 0511}, 017 (2005)
  [arXiv:hep-th/0412322].
  %%CITATION = HEP-TH 0412322;%%
  
%\cite{Aspinwall:2006yk}
\bibitem{Aspinwall:2006yk}
  P.~S.~Aspinwall, A.~Maloney and A.~Simons,
  ``Black hole entropy, marginal stability and mirror symmetry,''
  [arXiv:hep-th/0610033].
  %%CITATION = HEP-TH 0610033;%%

\bibitem{Tian:metrics}
G.~Tian,
``On a set of polarized K\"ahler metrics on algebraic
  manifolds'',
J.\ Diff.\ Geom.\ {\bf 32} (No. 1) 99--130 (1990).

\bibitem{Donaldson1}
S.~K. Donaldson,
``Scalar curvature and projective embeddings. I'',
J.\ Diff.\ Geom.\ {\bf 59} (No. 3) 479--522 (2001).

\bibitem{Donaldson:numeric}
S.~K. Donaldson,
``Some numerical results in complex differential geometry'',
[arXiv:math.DG/0512625].

%\cite{Jacobson:1995ab}
\bibitem{Jacobson:1995ab}
  T.~Jacobson,
  ``Thermodynamics of space-time: The Einstein equation of state,''
  Phys.\ Rev.\ Lett.\  {\bf 75}, 1260 (1995)
  [arXiv:gr-qc/9504004].
  %%CITATION = PRLTA,75,1260;%%
 
%\cite{Eling:2006aw}
\bibitem{Eling:2006aw}
  C.~Eling, R.~Guedens and T.~Jacobson,
  ``Non-equilibrium Thermodynamics of Spacetime,''
  Phys.\ Rev.\ Lett.\  {\bf 96}, 121301 (2006)
  [arXiv:gr-qc/0602001].
  %%CITATION = PRLTA,96,121301;%%

%\cite{Maldacena:1997de}
\bibitem{Maldacena:1997de}
  J.~M.~Maldacena, A.~Strominger and E.~Witten,
  ``Black hole entropy in M-theory,''
  JHEP {\bf 9712}, 002 (1997)
  [arXiv:hep-th/9711053].

%\cite{deBoer:2006vg}
\bibitem{deBoer:2006vg}
  J.~de Boer, M.~C.~N.~Cheng, R.~Dijkgraaf, J.~Manschot and E.~Verlinde,
  ``A farey tail for attractor black holes,''
  JHEP {\bf 0611}, 024 (2006)
  [arXiv:hep-th/0608059].
  %%CITATION = JHEPA,0611,024;%%

%\cite{Bachas:1999um}
\bibitem{Bachas:1999um}
  C.~P.~Bachas, P.~Bain and M.~B.~Green,
  ``Curvature terms in D-brane actions and their M-theory origin,''
  JHEP {\bf 9905} 011, (1999)
  [arXiv:hep-th/9903210].
  %%CITATION = JHEPA,9905,011;%%
  
%\cite{Green:1997di}
\bibitem{Green:1997di}
  M.~B.~Green and P.~Vanhove,
  ``D-instantons, strings and M-theory,''
  Phys.\ Lett.\  B {\bf 408}, 122 (1997)
  [arXiv:hep-th/9704145].
  %%CITATION = PHLTA,B408,122;%%

 %\cite{Marino:1999af}
\bibitem{Marino:1999af}
  M.~Marino, R.~Minasian, G.~W.~Moore and A.~Strominger,
  ``Nonlinear instantons from supersymmetric p-branes,''
  JHEP {\bf 0001}, 005 (2000)
  [arXiv:hep-th/9911206].

\bibitem{Balasubramanian:1996rx}
  V.~Balasubramanian and F.~Larsen,
  ``On D-Branes and Black Holes in Four Dimensions,''
  Nucl.\ Phys.\  B {\bf 478}, 199 (1996)
  [arXiv:hep-th/9604189].

\bibitem{Myers:1999ps}
  R.~C.~Myers,
  ``Dielectric-branes,''
  JHEP {\bf 9912}, 022 (1999)
  [arXiv:hep-th/9910053].

%\cite{deWit:1988ig}
\bibitem{deWit:1988ig}
  B.~de Wit, J.~Hoppe and H.~Nicolai,
  ``On the quantum mechanics of supermembranes,''
  Nucl.\ Phys.\  B {\bf 305}, 545 (1988).

%\cite{Banks:1996vh}
\bibitem{Banks:1996vh}
  T.~Banks, W.~Fischler, S.~H.~Shenker and L.~Susskind,
  ``M theory as a matrix model: A conjecture,''
  Phys.\ Rev.\  D {\bf 55}, 5112 (1997)
  [arXiv:hep-th/9610043].

%\cite{Iqbal:2003ds}
\bibitem{Iqbal:2003ds}
  A.~Iqbal, N.~Nekrasov, A.~Okounkov and C.~Vafa,
  ``Quantum foam and topological strings,''
  [arXiv:hep-th/0312022].

\bibitem{DKLR}
  M.~R.~Douglas, R.~L.~Karp, S.~Lukic and R.~Reinbacher,
   ``Numerical solution to the hermitian Yang-Mills equation on the Fermat
  quintic,''
  [arXiv:hep-th/0606261].
  %%CITATION = HEP-TH 0606261;%%
  
%\cite{Douglas:2008pz}
\bibitem{Douglas:2008pz}
  M.~R.~Douglas and S.~Klevtsov,
  ``Bergman Kernel from Path Integral,''
  arXiv:0808.2451 [hep-th].
  %%CITATION = ARXIV:0808.2451;%%
  
%\cite{Denef:2007vg}
\bibitem{Denef:2007vg}
  F.~Denef and G.~W.~Moore,
  ``Split states, entropy enigmas, holes and halos,''
  [arXiv:hep-th/0702146].
  
\bibitem{Zelditch:Szego}
S.~Zelditch,
``Szeg\"o kernels and a theorem of Tian,
Internat.\ Math.\ Res.\ Notices  (No. 6) 317--331 (1998).

\bibitem{Catlin}
D.~Catlin,
``The Bergman kernel and a theorem of Tian'', {\it Analysis and geometry in several complex variables (Katata,
  1997)'}, Trends Math., p. 1--23, {\it Birkh\"auser Boston, Boston} (1999).

\bibitem{LuZhiqin}
Z.~Lu,
``On the lower order terms of the asymptotic expansion of
  Tian-Yau-Zelditch'',
Amer.\ J.\ Math.\ {\bf 122} (No. 2) 235--273 (2000).

%\cite{Neitzke:2004ni}
\bibitem{Neitzke:2004ni}
  A.~Neitzke and C.~Vafa,
  ``Topological strings and their physical applications,''
  [arXiv:hep-th/0410178].

%\cite{Marino:2004eq}
\bibitem{Marino:2004eq}
  M.~Marino,
  ``Les Houches lectures on matrix models and topological strings,''
  [arXiv:hep-th/0410165].

\end{thebibliography}
\end{document}